
\input phyzzx

\def\df{\varphi}

\def\bg{{\hat g}}
\def\bgr{{\rm e}^{2\rho} {\hat g}}
\def\bgs{{\rm e}^{2\sigma} {\hat g}}
\def\blap{{\hat \Delta}}
\def\bcv{{\hat R}}
\def\vr{\delta \rho}

\def\vg{\delta g}
\def\vp{\delta \df}
\def\vf{\delta f}
\def\vx{\delta \xi}
\def\dg{\hbox{$\sqrt{-g}$}}
\def\dbg{\hbox{$\sqrt{-{\hat g}}$}}
\def\e{{\rm e}}
\def\det{{\rm det}}
\def\vepsi{\varepsilon}
\def\pp{\prime}
\def\pd{\partial}

\REF\hwa{S. Hawking, Comm. Math. Phys. {\bf 43} (1975) 199.}
\REF\hwb{S. Hawking, Phys. Rev. {\bf D14} (1976) 2460;
         J. Preskill, {\it Do Black Holes Destroy Information?}, preprint
         CALT-68-1819, reference therein.}
\REF\dfu{P. Davies, S. Fulling and W. Unruh, Phys. Rev. {\bf D13} (1976) 2720.}
\REF\cghs{C. Callan, S. Giddings, J. Harvey and A. Strominger, Phys. Rev.
          {\bf D45} (1992) R1005.}
\REF\rst{J. Russo, L. Susskind and L. Thorlacius, Phys. Lett. {\bf B292}
         (1992) 13;
         T. Banks, A. Dabholkar, M. Douglas and M. O'Loughlin, Phys. Rev.
         {\bf D45} (1992) 3607.}
\REF\w{S. Hawking, Phys. Rev. Lett. {\bf 69} (1992) 406;
       L. Susskind and L. Thorlacius, Nucl. Phys. {\bf B382} (1992) 123;
       B. Birnir, S. Giddings, J. Harvey and A. Strominger, Phys. Rev.
       {\bf D46} (1992) 638.}
\REF\h{K. Hamada, {\it Quantum Theory of Dilaton Gravity in 1+1 Dimensions},
       \break preprint UT-Komaba 92-7;
       {\it Gravitational Collapse in 1+1 Dimensions and Quantum Gravity},
        preprint UT-Komaba 92-9.}
\REF\str{S. deAlwis, Phys. Lett. {\bf B289} (1992) 278;
        A. Bilal and C. Callan, {\it Liouville Models of Black Hole
        Evaporation}, preprint PUPT-1320;
        S. Giddings and A. Strominger, {\it Quantum Theories of Dilaton
        Gravity}, preprint UCSB-TH-92-28.}
\REF\rt{J. Russo and A. Tseytlin, Nucl. Phys. {\bf B382} (1992) 259;
        A. Mikovi\'c, Phys. Lett. {\bf B291} (1992) 19.}
\REF\os{S. Odintsov and I. Shapiro, Phys. Lett. {\bf B263} (1991) 183;
        {\it Perturbative analysis of two-dimensional quantum gravity},
        Int. J. Mod. Phys. {\bf A}, to appear.}
\REF\c{A. Chamseddine, Phys. Lett. {\bf B256} (1991) 379; {\bf B258} (1991)
       97; Nucl. Phys. {\bf B368} (1992) 98;
       I. Lichtzier and S. Odintsov, Mod. Phys. Lett. {\bf A6} (1991) 1953;
       T. Burwick and A. Chamseddine, {\bf B384} (1992) 411.}
\REF\dk{J. Distler and H. Kawai, Nucl. Phys. {\bf B321} (1989) 509;
        F. David, Mod. Phys. Lett. {\bf A3} (1988) 1651.}
\REF\fr{D. Friedan, In {\it Recent Advances in Field Theory and Statistical
        Mechanics} ({\it Les Houches, 1982}), edited by J. Zuber and
        R. Stora, North-Holland.}
\REF\al{O. Alvarez, Nucl. Phys. {\bf B216} (1983) 125;
        E. D'Hoker and D. Phong, Rev. Mod. Phys. {\bf 60} (1988) 917.}
\REF\no{S. Nojiri and I. Oda, Phys. Lett. {\bf B294} (1992) 317.}
\REF\mmr{See also R. Mann, M. Morris and S. Ross, {\it Properties of
         Asymptotically Flat Two-Dimensional Black Holes}, preprint WATPHYS
         TH-91/04.}
\REF\y{See also T. Yoneya, Prog. Theor. Phys. Suppl. {\bf 85} (1985) 256;
       Phys. Lett. {\bf B149} (1984) 111.}
\REF\ct{T. Curtright and C. Thorn, Phys. Rev. Lett. {\bf 48} (1982) 1309;
        E. D'Hoker, D. Freedman and R. Jackiw, Phys. Rev. {\bf D28} (1984)
        2583; T. Yoneya, Phys. Lett. {\bf B148} (1984) 111;
        N. Seiberg, Prog. Theor. Phys. Suppl. {\bf 102} (1990) 319.}
\REF\tih{P. Thomi, B. Isaak and P. Hajicek, Phys. Rev. {\bf D30} (1984) 1168;
         P. Hajicek, Phys. Rev. {\bf D30} (1984) 1178.}
\REF\his{W. Hiscock, Phys. Rev. {\bf D23} (1981) 2813.}
\REF\l{Numerical analyses appear in D. Lowe, {\it Semiclassical Approach to
       Black Hole Evaporation}, preprint PUPT-1340.}

\pubnum{UT-Komaba 92-14}
\date={December 1992}

\titlepage

\title{{\bf Quantum Gravity and Black Hole Dynamics \break
            in 1+1 Dimensions }}

\author{Ken-ji Hamada\footnote1{E-mail address:
        hamada@tkyvax.phys.s.u-tokyo.ac.jp} and
        Asato Tsuchiya\footnote2{E-mail address:
        tutiya@tkyvax.phys.s.u-tokyo.ac.jp}}

\address{Institute of Physics, University of Tokyo \break
         Komaba, Meguro-ku, Tokyo 153, Japan \break}

\centerline{\bf Abstract}

     We study the quantum theory of 1+1 dimensional dilaton
gravity, which is an interesting toy model of the black hole dynamics. The
functional measures are explicitly evaluated and the physical state conditions
corresponding to the Hamiltonian and the momentum constraints are derived.
It is pointed out that the constraints form the Virasoro algebra without
central charge.
In ADM formalism the measures are very ambiguous, but in our
formalism they are explicitly defined.
Then the new features which are not seen in ADM formalism come out.
A singularity appears at $\df^2 =\kappa (>0) $, where
$\kappa =(N-51/2)/12 $ and $ N$ is the number of matter fields. Behind the
singularity the quantum mechanical region $\kappa > \df^2 >0 $ extends, where
the sign of the kinetic term in the Hamiltonian constraint changes.
If $\kappa <0 $, the singularity disappears.
We discuss the quantum dynamics of black hole and then  give a suggestion for
the resolution of the information loss paradox.
We also argue the quantization of the spherically symmetric gravitational
system in 3+1 dimensions. In appendix the differences between the other
quantum dilaton gravities and ours are clarified and our status is stressed.

\endpage

\chapter{\bf Introduction}

   The quantum dynamics of the black hole is an important issue relating to
fundamental laws of physics both in cosmology and field theories.
Since the discovery of Hawking,\refmark{\hwa} many authors have
investigated whether the usual rules of quantum mechanics can be applied to
quantum black holes or not.\refmark{\hwb} Do black holes
really evaporate and, if it is true, are informations indeed lost?
No definite  argument has not been yet. To resolve these problems the gravity
also should be quantized.

   The black hole evaporation is caused by non-perturbative quantum effects.
Davies, Fulling and Unruh\refmark{\dfu} discussed the black hole dynamics
in two dimensional equivarent of
the Schwarzchild black hole and showed that the conformal anomaly induces
the emission of thermal radiation. This indicates that when we argue the black
hole dynamics we must carefully evaluate divergence properties of
quantum fields.

  As a quantization method of gravitation, Arnowitt-Deser-Misner (ADM)
formalism or Wheeler-DeWitt approach is well-known. There are, however,
some serious problems in ADM formalism, which are the issues of measures and
orderings. These are the most important points when we discuss quantum
field theories. As far as ignoring these effects we cannot say anymore beyond
WKB approximation. Anomalies cannot be derived from WKB approximation. Namely,
it is necessary to quantize the gravitation exactly when we discuss the
dynamics of black hole.

   Recently Callan, Giddings, Harvey and Strominger\refmark{\cghs} proposed
an interesting toy model of gravity in 1+1 dimensions.
It is called the dilaton gravity.
The model has interesting features similar to the spherically symmetric
gravitational system in 3+1 dimensions.
The essence of the black hole dynamics appears to be
included enough. Really in the semi-classical approximation the dynamics
can be discussed in completely parallel with the case of the spherically
symmetric black hole.
Furthermore they advanced the arguments so that the gravitational
back-reaction effects were included systematically by introducing the large
number of matter fields.\refmark{\cghs,\rst,\w}

   In this paper we develop the argument to the quantum gravity.\refmark{\h}
In Sect.2 we first define the quantum theory of the dilaton gravity and
clarify the differences from the other definitions\refmark{\str,\rt,\os,\c}
(see also appendix). Then our status is stressed.
We explicitly evaluate the contributions of measures of gravity part
and fix the diffeomorphism invariance completely in conformal gauge by using
the techniques developed in two dimensional
quantum gravity.\refmark{\dk,\fr,\al}
  In Sect.3 and 4 we derive the physical state conditions that
correspond to the Hamiltonian and the momentum constraints and discuss the
algebraic structure of them.
Then the new features which are not seen in ADM formalism come out.
A singularity appears at $\df^2 =\kappa (>0) $, where
$\kappa =(N-51/2)/12 $ and $ N$ is the number of matter fields. Behind the
singularity the quantum mechanical region $\kappa > \df^2 >0 $ extends, where
the sign of the kinetic term in the Hamiltonian constraint changes.
If $\kappa <0 $, the singularity disappears. The existence of the quantum
mechanical region gives a new insight when we discuss the dynamics of black
holes in Sect.5. We argue a possibility of gravitational tunneling and
give a suggestion for the resolution of the information loss paradox.
In Sect.6 we attempt to quantize the spherically symmetric gravitational
system in 3+1 dimensions. In this case some problems appear.

\chapter{\bf Quantum dilaton gravity}

     The theory of 1+1 dimensional dilaton gravity is defined by the
following action\footnote\dagger{Here we do not discuss the model coupled with
gauge fields, which is discussed in ref.15.}
$$
\eqalign{
       &  I(g,\df,f)=I_D (g,\df)+I_M (g,f) ~,               \cr
       &  I_D (g,\df)= {1 \over 2\pi} \int d^2 x \dg
               (R_g \df^2
                +4 g^{\alpha \beta} \partial_{\alpha} \df \partial_{\beta} \df
                +4 \lambda^2 \df^2 ) ~,        \cr
       &  I_M (g,f) = -{1 \over 4\pi} \sum^N_{j=1} \int d^2 x \dg
           g^{\alpha \beta} \partial_{\alpha} f_j \partial_{\beta} f_j ~, \cr
        } \eqno\eq
$$
where $\df ={\rm e}^{-\phi} $ is the dilaton field and $f_j $'s are $N$ matter
fields. $\lambda^2 $ is the cosmological constant.  $R_g $ is the
curvature of the metrics $g $. The classical equations of motion can be solved
exactly and one obtains, for instance, the black hole geometry
$$
        \df^2 = {\rm e}^{-2\rho} = {M \over \lambda} -\lambda^2 x^+ x^- ~,
          \qquad f_j =0 ,
        \eqno\eq
$$
where $g_{\alpha \beta}={\rm e}^{2\rho} \eta_{\alpha \beta} $,
$\eta_{\alpha \beta} =(-1,1) $ and $x^{\pm} =x^0 \pm x^1 $. $M $ is the mass
of the black hole. More interesting geometry is the gravitational
collapse.\refmark{\mmr} It is given by
$$
       \df^2 =\e^{-2\rho} =
         -{M \over \lambda x^+_0 } (x^+ -x^+_0 ) \vartheta (x^+ -x^+_0 )
            -\lambda^2 x^+ x^-  ~,
       \eqno\eq
$$
where $\vartheta $ is the step function. The infalling matter flux is given
by the shock wave along the line $x^+ =x^+_0 $
$$
     {1 \over 2} \sum^N_{j=1} \pd_+ f_j \pd_+ f_j
        = {M \over \lambda x^+_0 } \delta (x^+ -x^+_0 )  ~.
    \eqno\eq
$$

    The quantum theory of the dilaton gravity is defined by
$$
    Z = \int {D_g(g) D_g(\df) D_g(f) \over {\rm Vol(Diff.)} }
           {\rm e}^{iI(g,\df,f)} ~,
     \eqno\eq
$$
where ${\rm Vol(Diff.)} $ is the gauge volume. The functional measures are
defined from the following norms
$$
\eqalign{
    &  < \vg, \vg >_g =
           \int d^2 x \dg g^{\alpha \beta} g^{\gamma \delta}
            (\vg_{\alpha \gamma} \vg_{\beta \delta}
                + u \vg_{\alpha\beta} \vg_{\gamma\delta}) ~,        \cr
    &  < \vp, \vp>_g =
           \int d^2 x \dg \vp \vp  ~,                           \cr
    &  < \vf_j , \vf_j >_g =
           \int d^2 x \dg \vf_j \vf_j
             \qquad (j=1, \cdots N)  ~,                       \cr
          }   \eqno\eq
$$
where $u>-1/2 $.\footnote\flat{Then the measure (2.13) becomes positive
definite.}
 The integration range of $\df $ is the whole real values. Physically we should
restrict the values of $\df $ within the non-negative values. However, since
the action (2.1) is invariant under the change $\df \rightarrow -\df $,
it seems that our definition is meaningful enough when we discuss the quantum
dynamics of black holes.

   The several authors discuss the other type of quantum
theory.\refmark{\str,\rt,\os,\c,\y} If we carry out the field transformation
$$
      \chi =\df^2 ~, \qquad h =\e^{2\omega}g ~, \quad
         \omega = {1 \over 2} \log \chi -{1 \over 2} \chi  ~,
      \eqno\eq
$$
the classical dilaton action becomes\refmark{\rt,\os,\y}
$$
     I_D (h, \chi)= {1 \over 2\pi}
            \int d^2 x \hbox{$\sqrt{-h}$}
               ( R_h \chi + h^{\alpha \beta}
               \partial_{\alpha} \chi \partial_{\beta} \chi
                +4 \lambda^2 \e^{\chi} ) ~.
      \eqno\eq
$$
The matter action does not change under this transformation. Then the measures
are defined for the fields $\chi $ and $h $ instead of
$\df $ and $g $. This is a definition of quantum gravity,
but this definition has a demerit.
The theory does not have the $Z_2 $ symmetry under the change
$\chi \rightarrow -\chi $ so that the restriction to $\chi \geq 0 $ seems to
be crucial. Thus it is not suited for discussing the quantum dynamics of black
holes. If one ignores the restriction, the quantum theory  becomes
very simple. It reduces to a free-like field theory, which means that the
short distance behavior becomes that of the usual free field in two
dimensions. The quantization of this theory is discussed in appendix.
The quantum theories of ref.8 are very similar to this one.
On the other hand the quantum theory (2.5) has quite different features as
discussed below.  Really it does not become a free-like theory.

   Let us first discuss the measure of the metrics. We decompose the metrics
into a conformal factor $\rho $ and a background metric $\bg $ as
$ g = \bgr $. This is the conformal gauge-fixing condition adopted here.
The change in the
metric is given by the change in the conformal factor $\vr $
and the change under a
diffeomorphism $ \vx_{\alpha}$ as
$$
\eqalign{
          \vg_{\alpha \beta}
             & = 2\vr g_{\alpha \beta}
               + \nabla_{\alpha} \vx_{\beta}
               + \nabla_{\beta} \vx_{\alpha}           \cr
             & = 2 \vr^{\pp} g_{\alpha \beta}
               + (P_1 \vx )_{\alpha \beta} ~,          \cr
         }    \eqno\eq
$$
where
$$
       \vr^{\pp} = \vr + {1 \over 2} \nabla^{\gamma} \vx_{\gamma} ~,
         \qquad
       (P_1 \vx)_{\alpha \beta} = \nabla_{\alpha} \vx_{\beta}
                  + \nabla_{\beta} \vx_{\alpha}
                  - g_{\alpha \beta} \nabla^{\gamma} \vx_{\gamma} ~.
      \eqno\eq
$$
The variations $ \vr^{\pp} g_{\alpha\beta} $ and $ (P_1 \vx)_{\alpha\beta} $
are orthogonal in the functional space defined by the norms (2.6). Therefore
the measure over metrics can be decomposed as
$$
\eqalign{
         D_g (g) & = D_g (\rho^{\pp}) D_g (P_1 \xi)    \cr
                 & = D_g (\rho)  D_g (\xi_{\alpha}) \det_g P_1  ~.    \cr
        }   \eqno\eq
$$
The functional integration over $\xi_{\alpha} $ cancels out the gauge volume.
The Jacobian $\det_g P_1 $ can be represented by  the functional integral
over the ghosts $b, c$. Thus the partition function (2.5) becomes
$$
     Z = \int D_g (\rho) D_g (\df) D_g (f) D_g (b) D_g (c)
           \exp \bigl[ iI_D (g,\df)+iI_M (g,f)+iI_{gh}(g,b,c) \bigr]  ~,
      \eqno\eq
$$
where  $I_{gh} $ is the well-known ghost action (see for
example  ref.13). The measure $D_g (\rho) $ is defined from the norm (2.6) by
$$
    <\vr, \vr>_g = \int d^2 x \hbox{$\sqrt{-g} $} (\vr)^2
                 = \int d^2 x \hbox{$\sqrt{-\bg} $} \e^{2\rho} (\vr)^2 ~.
     \eqno\eq
$$

   This is not the end of the story.  The expression (2.12) has serious
problems. The measure (2.13) is not invariant under the local shift
$\rho \rightarrow \rho +\epsilon $ and also the measures of the fields
$\df, f, b$
and $c $ explicitly depend on the dynamical variable $g =\bgr $. This is quite
inconvenient because we must pick up contributions from the measures when
the conformal factor $\rho $ is integrated. So we will rewrite  the measures
on $ g $ into more convenient ones  defined on the background metric $\bg $.

    First we rewrite the measures of the dilaton, the matter and
the ghost fields into the convenient ones. For the measures of the matter and
the ghost fields it is realized by using the well-known transformation
property (see for example ref.14)
$$
     D_{\bgr}(f) D_{\bgr}(b) D_{\bgr}(c)
          = \exp \biggl[ i{N-26 \over 12\pi} S_L (\rho, \bg) \biggr]
            D_{\bg}(f) D_{\bg}(b) D_{\bg}(c)  ~,
     \eqno\eq
$$
where $S_L (\rho, \bg) $ is what is called the Liouville action defined by
$$
        S_L (\rho, \bg) = {1 \over 2}
                  \int d^2 x \dbg ( \bg^{\alpha \beta}
                   \partial_{\alpha} \rho \partial_{\beta} \rho
                   + \bcv \rho ) ~.
          \eqno\eq
$$
Note that the actions of the matter and the ghost fields are invariant
under the Weyl rescalings, or $ I_M (g, f) =I_M (\bg, f) $ and
$I_{gh}(g,b,c)=I_{gh}(\bg,b,c) $.

   For the measure of the dilaton field the following relation is realized,
$$
        \int D_{\bgr} (\df) \e^{iI_D (\bgr,\df)}
           = \exp \biggl[ i{c_{\df} \over 12\pi} S_L (\rho, \bg) \biggr]
               \int D_{\bg}(\df) \e^{iI_D (\bgr, \df)}
        \eqno\eq
$$
with $c_{\df} = -1/2 $. A notable point is that the dilaton action $I_D $ is
not invariant under the Weyl rescalings. Pay attention to the $\rho
$-dependence of each side of (2.16). This expression is proved by comparing the
$\rho $-dependence of the functional integrations of each side.
The l.h.s. gives the determinant
$$
     \int D_g (\df) \e^{iI_D (g, \df)}
            = L [\det_g D ]^{-1/2} ~, \qquad g=\bgr ~,
      \eqno\eq
$$
where the operator $D $ is defined by
$$
\eqalign{
         D & = \Delta_g + {1 \over 4} R_g + \lambda^2    \cr
           & = \e^{-2\rho} \blap + {1 \over 4} \e^{-2\rho}
                (\bcv +2 \blap \rho ) + \lambda^2        \cr
        } \eqno\eq
$$
and $L $ is a constant factor and $\Delta $ is the Laplacian defined by
$- \nabla^{\alpha} \nabla_{\alpha}$. The functional integration of r.h.s.
gives the determinant
$$
     \int D_{\bg} (\df) \e^{iI_D (\bgr, \df)}
            = L [\det_{\bg} {\hat D} ]^{-1/2} ~,
      \eqno\eq
$$
where ${\hat D} $ is defined by
$$
     {\hat D} \equiv \e^{2\rho} D
              = \blap + {1 \over 4}(\bcv + 2\blap \rho)
                  + \lambda^2 \e^{2\rho} ~.
      \eqno\eq
$$
The determinants (2.17) and (2.19) can be evaluated by using the heat-kernel
method. Here we want to know only the difference between them. Paying
attention to the $\rho $-dependence, we get the simple relation
$$
\eqalign{
          &  \delta_{\rho}  {\rm log} \det_g D
              -  \delta_{\rho} {\rm log} \det_{\bg} {\hat D}      \cr
          & = -2 Tr(\vr \e^{-i\vepsi D} )                      \cr
          & = \delta_{\rho} \biggl[   - i{c_{\df} \over 12\pi }
                   \int d^2 x \dbg (\rho \blap \rho +\bcv \rho)
                   + \Lambda \int d^2 x \dbg \e^{2\rho}
                    \biggr]  ~,                                    \cr
        }   \eqno\eq
$$
where $\vepsi $ is a infinitesimal parameter to regularize divergences.
$\Lambda $ is the divergent constant
${1 \over 4\pi} ( -{1 \over \vepsi} +i\lambda^2 )$, which is renormalized
to zero by introducing a bare term $\mu_0 \int d^2 x \sqrt{-g} $ and adjusting
the bare constant $\mu_0 $ properly. The details of the calculation appear in
ref.7. From eq.(2.21) we obtain the expression (2.16).

   From the expression (2.14) and (2.16) we get
$$
\eqalign{
   Z =  \int D_{\bgr}(\rho) & D_{\bg}(\df) D_{\bg}(f) D_{\bg}(b) D_{\bg}(c)
          \exp \biggl[ i {c_{\df}+N-26 \over 12\pi} S_L (\rho, \bg)      \cr
          &  + iI_D (\e^{2\rho} \bg,\df) +iI_M (\bg,f)
               +iI_{gh}(\bg,b,c) \biggr] ~.                          \cr
        }   \eqno\eq
$$

    Next we rewrite the measure of $\rho $. According to the procedure of
David-Distler-Kawai (DDK),\refmark{\dk} we assume the following relation
$$
       D_{\bgr} (\rho) = D_{\bg} (\rho)
            \exp \biggl[ i {A \over 12\pi} S_L (\rho, \bg) \biggr] ~.
       \eqno\eq
$$
Note that the measure $D_{\bg} (\rho) $ is invariant under the local shift of
$\rho $. The parameter $A $ is determined by the consistency. Since the
original theory depends only on the metrics $g= \bgr $, the theory should be
invariant under the simultaneous shifts
$$
        \rho \rightarrow \rho - \sigma ~, \qquad \bg \rightarrow \bgs ~.
          \eqno\eq
$$
This requirement leads to $A=1 $. The exact proof is given in ref.7.
Finally we get the expression
$$
      Z = \int D_{\bg} (\Phi) \e^{i{\hat I}(\bg,\Phi)} ~,
         \eqno\eq
$$
where $\Phi $ denotes the fields $\rho, \df, f, b $ and $c $. ${\hat I} $ is
the gauge-fixed action
$$
\eqalign{
   {\hat I} = {1 \over 2\pi} & \int d^2 x \dbg \Bigl[
    4{\hat g}^{\alpha \beta} \partial_{\alpha} \df \partial_{\beta} \df
    +4{\hat g}^{\alpha \beta} \df \partial_{\alpha} \df \partial_{\beta} \rho
       + \bcv \df^2 +4 \lambda^2 \df^2 \e^{2\rho}          \cr
      & + \kappa (
         {\hat g}^{\alpha \beta} \partial_{\alpha} \rho \partial_{\beta} \rho
         + \bcv \rho )
       -{1 \over 2} \sum^N_{j=1}
         {\hat g}^{\alpha \beta} \partial_{\alpha} f_j \partial_{\beta} f_j
       \Bigr] +I_{gh}(\bg,b,c)                             \cr
        }  \eqno\eq
$$
with
$$
      \kappa ={1 \over 12}(1+c_{\df}+N-26)={N-51/2 \over 12} ~.
        \eqno\eq
$$

    Closing this section there are some remarks. We showed that the theory
(which includes the
measures) is invariant under the simultaneous shifts (2.24). Furthermore  the
measure $D_{\bg} (\rho)$ is invariant under the local shift of $\rho $. So the
theory is invariant under conformal changes of the background metric $\bg $:
$\bg \rightarrow \bgs$. More explicitly the Liouville-dilaton part is
transformed as
$$
\eqalign{
        & \int D_{\bgs} (\rho) D_{\bgs}(\df)
              \exp \biggl[ i{\kappa \over \pi} S_L (\rho, \bgs)
                      +iI_D (\e^{2\rho}\bgs,\df) \biggr]        \cr
        & = \int D_{\bgs} (\rho) D_{\bgs}(\df)
              \exp \biggl[ i{\kappa \over \pi} S_L (\rho-\sigma, \bgs)
                      +iI_D (\bgr,\df) \biggr]        \cr
        & = \exp \biggl[ i{1+c_{\df} \over 12\pi}S_L(\sigma,\bg) \biggr]
            \int D_{\bg} (\rho) D_{\bg}(\df)
              \exp \biggl[ i{\kappa \over \pi} S_L (\rho-\sigma, \bgs)
                      +iI_D (\bgr,\df) \biggr]        \cr
        & = \exp \biggl[ -i{N-26 \over 12\pi}S_L(\sigma,\bg) \biggr]
            \int D_{\bg} (\rho) D_{\bg}(\df)
              \exp \biggl[ i{\kappa \over \pi} S_L (\rho, \bg)
                      +iI_D (\bgr,\df) \biggr]  ~,      \cr
       }   \eqno\eq
$$
where in the last equality we use the relation for the Liouville action
$$
       S_L (\rho-\sigma, \bgs) = S_L (\rho,\bg)-S_L (\sigma,\bg) ~.
        \eqno\eq
$$
The extra Liouville action $-i{N-26 \over 12\pi} S_L(\sigma,\bg)$ cancels
out with that induced from the measures of the matter and ghost fields
(see eq.(2.14)) so that the partition function is invariant
under the conformal change of $\bg $. This invariance is quite reasonable
because the background metric $\bg $ is very artificial. The theory should be
independent of how to choose the background metric.

    Here there is a question whether the theory (2.25) is regarded as a kind
of conformal field theory (CFT) on $\bg $ or not.  The usual definition of CFT
is that the action is invariant under the conformal transformation. According
to this definition the Liouville theory is not CFT. However, the Liouville
action satisfies the relation (2.29), which means that the Poisson brackets of
the classical energy-momentum tensor satisfy the Virasoro algebra with central
extention $-12\kappa $. Furthermore, as shown in ref.18, the quantum Liouville
theory realizes the Virasoro algebra with central charge $c_{\rho}=1-12\kappa$
(which is easily derived by ignoring the dilaton field in eq.(2.28)).
Thus the Liouville theory is considered as a kind of CFT.
In the theory (2.25), we must treat the fields $\rho $ and $\df $ in pairs
because the theory has the derivative coupling of the \lq\lq third order"
of fields. The equation (2.28) says that the Liouville-dilaton part of the
quantum energy-momentum tensor satisfies the Virasoro algebra with central
extension $c_{\rho \df} =1+c_{\df}-12\kappa =26-N $. In general CFT is
described by a set of free fields, while the theory
(2.25) has the non-trivial coupling and is not free-like so that it is quite
different from usual CFT.  The Virasoro structrue of this theory is realized
in the non-trivial way, which is discussed in Sect.4.

   The second remark is that the partition function is a scalar.
This is manifest in the definition (2.5). After rewriting the partition
function into the expression (2.25), however, this invariance is hidden.
It is instructive to
show that the partition function is really scalar. The Liouville field $\rho $
is transformed as
$$
      \rho^{\pp} (x^{\pp})=\rho (x) -\gamma (x) ~, \qquad
        \gamma (x) = {1 \over 2} \log
                     \biggl\vert {\pd x^{\pp} \over \pd x} \biggr\vert^2  ~,
      \eqno\eq
$$
where we only consider the conformal coordinate transformation
$x^{\pm \pp} =x^{\pm \pp}(x^{\pm}) $ to preserve the conformal gauge and use
the notation $\vert x \vert^2 = x^+ x^- $. On the other hand
the background metric is not transformed: $\bg^{\pp}(x^{\pp}) =\bg (x) $.
It is natural because the background metric is not dynamical.
Therefore the gauge-fixed action is transformed as
$$
      {\hat I}^{\pp} ={\hat I}-{\kappa \over \pi} S_L (\gamma, \bg) ~,
        \eqno\eq
$$
where note that $R_g $ is a scalar, but $\bcv $ is transformed as
$\bcv^{\pp} =\vert {\pd x \over \pd x^{\pp}}
\vert^2 (\bcv +2\blap \gamma) $. The measures defined on $\bg $ are also
non-invariant under the coordinate transformation. The extra Liouville term
$S_L (\gamma,\bg) $ cancels out with that coming from the measures so that
the partition function is invariant. By replacing $\gamma $ with the conformal
change $\sigma $, it is seen that the invariance under the
conformal change of $\bg $ after all guarantees the
invariance under the coordinate transformation.

\chapter{{\bf Physical state conditions}}

     Now we carry out the canonical quantization of the gauge-fixed 1+1
dimensional dilaton gravity. As mentioned in Sect.2 the theory
should be independent of how to choose the background metric $\bg $. Thus
the variation of the partition function with respect to $\bg $ vanishes
$$
    0= {\delta Z \over \delta {\hat g}^{\alpha \beta} }
     = \int D_{{\hat g}} (\Phi)
           i {\delta {\hat I} \over \delta {\hat g}^{\alpha \beta} }
            {\rm e}^{i{\hat I}({\hat g},\Phi)}
      + \int {\delta D_{{\hat g}} (\Phi) \over \delta {\hat g}^{\alpha \beta}}
            {\rm e}^{i{\hat I}({\hat g},\Phi)} ~.
    \eqno\eq
$$
The first term of r.h.s. is nothing but
$< i{\delta {\hat I} \over \delta {\hat g}^{\alpha \beta} } >_{{\hat g}} $.
The second term picks up an anomalous contribution.
But if we choose the Minkowski background ${\hat g} =\eta $,
this contribution vanishes. So it is convenient to choose the Minkowski
background metric. Then the physical state conditions are
$$
       \langle {\delta {\hat I} \over \delta \bg^{\alpha \beta}}
            \rangle_{\bg =\eta} =0
           \eqno\eq
$$
or
$$
      < {\hat T}_{00} >_{\bg=\eta} = <{\hat T}_{01} >_{\bg=\eta} =0 ~,
            \eqno\eq
$$
where the energy-momentum tensor ${\hat T}_{\alpha \beta} $ is defined by
${\hat T}_{\alpha \beta}=-{2 \over \dbg}{\delta {\hat I} \over \delta
\bg^{\alpha\beta}} \vert_{\bg=\eta}$.
The condition for ${\hat T}_{11} $ reduces to the
one for $ {\hat T}_{00} $ by using the $\rho $-equation of motion.
Furthermore we restrict the physical state to the one which satisfies the
condition $<{\hat T}^{gh}_{\alpha\beta}>_{\bg=\eta}=0$ because the ghost flux
should vanish in the flat space time.

   Since the functional measures are defined on the Minkowski background
metric, we can set up the canonical commutation relations as usual.
The conjugate momentums for $\rho, \df $ and $f_j $ are given by
$$
\eqalign{
     \Pi_{\rho}&= -{\kappa \over \pi} {\dot \rho}
                       -{2 \over \pi} \df {\dot \df} ~,        \cr
     \Pi_{\df} &= -{4 \over \pi} {\dot \df}
                       -{2 \over \pi} \df {\dot \rho}  ~,      \cr
     \Pi_{f_j} &= {1 \over 2\pi}  {\dot f}_j  ~,               \cr
        } \eqno\eq
$$
where the dot stands for the derivative with respect to the
time coordinate. Then the physical state conditions (3.3) can be expressed as
$$
\eqalign{
       \biggl[ {\pi/2 \over \df^2 -\kappa}
           & \Bigl( \Pi^2_{\rho} -\df \Pi_{\df} \Pi_{\rho}
                  + {\kappa \over 4} \Pi^2_{\df} \Bigr)
           + {2 \over \pi}  \bigl(  \df \df^{\prime\prime}
                 -\df \df^{\prime} \rho^{\prime}
                 -\lambda^2 \df^2 \e^{2\rho} \bigr)                  \cr
           & -{\kappa \over 2\pi}
               \bigl( \rho^{\prime 2} -2\rho^{\prime\prime} \bigr)
           +\sum^N_{j=1}  \Bigl( \pi \Pi^2_{f_j}
                   + {1 \over 4\pi} f^{\prime 2}_j \Bigr)
                     \biggr]  \Psi =0                                 \cr
         }    \eqno\eq
$$
and
$$
      \bigl(  \rho^{\pp} \Pi_{\rho} - \Pi^{\pp}_{\rho}
               + \df^{\pp} \Pi_{\df} + \sum^N_{j=1} \Pi_{f_j} f^{\pp}_j
                   \bigr) \Psi =0    ~,
       \eqno\eq
$$
where $\kappa $ is defined by eq.(2.27). $\Psi $ is a physical state.
The prime stands for the derivative with respect to the space coordinate.

  Here we have two remarks. The first is that the fields $\rho $ and $\df $
are dynamical variables so that it is significant to consider the equations
of motion of $\rho $ and $\df $. But $\bg $ is not dynamical. So we should
not regard the physical state conditions as the equations of motion of $\bg$.
The conditions come from the symmetry of the  theory. In this point
of view the conditions indeed correspond to the constraints.
Therefore we call eqs.(3.5) and (3.6) the Hamiltonian and the momentum
constraints respectively. These are the modified versions of the
Wheeler-DeWitt equations.\footnote\sharp{The usual Wheeler-DeWitt equations
are derived, for example, in ref.19, where the spherically symmetric
gravitational system in 3+1 dimensions is discussed. Application to the
1+1 dimensional dilaton gravity is straightforward.}.

  The second remark is that the energy-momentum tensor
${\hat T}_{\alpha\beta} $ is transformed as non-tensor
because the Liouville
field $\rho $ is transformed as (2.30) for the conformal coordinate
transformation $x^{\pm\pp}=x^{\pm\pp}(x^{\pm}) $. In the light-cone
coordinate we get
$$
\eqalign{
         {\hat T}^{\pp}_{\pm\pm}(x^{\pp})
            & = \biggl( {\pd x^{\pm} \over \pd x^{\pm \pp} } \biggr)^2
              \bigl( {\hat T}_{\pm\pm}(x)
                    + {\kappa \over \pi} t_{\pm}(x) \bigr)  ~,   \cr
         {\hat T}^{\pp}_{+-}(x^{\pp})
            & = \biggl\vert {\pd x \over \pd x^{\pp}} \biggr\vert^2
              {\hat T}_{+-}(x)   ~,                                \cr
        }  \eqno\eq
$$
where $t_{\pm}(x)$ is the Schwarzian derivative
$$
     t_{\pm}(x) =  {\pd \gamma(x) \over \pd x^{\pm}}
                        {\pd \gamma(x) \over \pd x^{\pm}}
                         -{\pd^2 \gamma(x) \over \pd x^{\pm 2}} ~, \qquad
     \gamma(x)={1 \over 2} \log \biggl\vert
                          {\pd x^{\pp} \over \pd x}
                             \biggr\vert^2 ~.
   \eqno\eq
$$
Therefore the physical state conditions (3.5-6) correspond to the
case of $t_{\pm} =0 $. To determine what coordinate system corresponds to
this case is a physical requirement.
It is natural that the coordinate system which is joined to the Minkowski
space time (asymptotically) is considered as the coordinate system with
$t_{\pm} =0$.

   If $ \kappa > 0 $, there is a singularity at finite $\df^2 =\kappa $.  The
region $\df^2 >\kappa $ is the classically allowed
region,\footnote\natural{Here ${\hat I}$ is considered as a classical action}
whereas the region $\kappa > \df^2 >0 $ is called the Liouville region, where
the sign of the kinetic term of the Hamiltonian constraint changes.
This is the classically
forbidden region. The existence of the Liouville region is interesting. There
may be some possibility of gravitational tunneling through this region.
If $\kappa <0 $, the situation drastically changes. In this
case the singularity  disappears.

\chapter{{\bf On Virasoro algebra in quantum dilaton gravity}}

    The constraints should form the closed algebra without central extension.
We first discuss the Poisson brackets between the constraints. The Poisson
brackets are defined by
$$
     [\rho(x), \Pi_{\rho}(y)]_{P.B.}=\delta(x-y) ~,
      \qquad [\df(x), \Pi_{\df}(y)]_{P.B.}=\delta(x-y) ~.
         \eqno\eq
$$
Here we concentrate on the Liouville-dilaton part. Then the Poisson brackets
become
$$
\eqalign{
         [H^{\rho\df}(x), H^{\rho\df}(y)]_{P.B.}
           & = 2 P^{\rho\df}(x) \delta^{\pp}(x-y)
                  + P^{\rho\df \pp}(x) \delta(x-y) ~,       \cr
         [P^{\rho\df}(x), P^{\rho\df}(y)]_{P.B.}
           & = 2 P^{\rho\df}(x) \delta^{\pp}(x-y)
                  + P^{\rho\df \pp}(x) \delta(x-y) ~,       \cr
         }  \eqno\eq
$$
and
$$
\eqalign{
         & [H^{\rho\df}(x),  P^{\rho\df}(y)]_{P.B.}
                 = [P^{\rho\df}(x), H^{\rho\df}(y)]_{P.B.}            \cr
          & =  2 H^{\rho\df}(x) \delta^{\pp}(x-y)
                 + H^{\rho\df \pp}(x) \delta(x-y)
                 + {\kappa \over \pi} \delta^{\pp\pp\pp}(x-y) ~,   \cr
         }  \eqno\eq
$$
where
$$
        H^{\rho\df}(x) = {\hat T}^{\rho\df}_{00}(x) ~, \qquad
        P^{\rho\df}(x) = {\hat T}^{\rho\df}_{01}(x) ~.
          \eqno\eq
$$
The Poisson brackets of the matter and the ghost parts are the same as the
expression above without the central extension.

   The central extension of the Poisson bracket (4.3) refrects that the
gauge-fixed action ${\hat I} $ is not invariant under the coordinate
transformation and transformed as eq.(2.31). The extra Liouville action of
eq.(2.31) indeed corresponds to the central extension of the Poisson bracket.

   The results in the path integral show that the conformal invariance is
recovered by the quantum corrections which come from the measures defined on
$\bg $. In terms of the operator formalism it means that, if we replace the
Poisson brackets with the commutators and define the normal ordering properly,
the central term of the Poisson bracket is canceled out completely.

   What is the proper normal ordering consistent to the path integral
results? For the matter and the ghost fields we can adopt the free field
normal ordering, but for the Liouville and the dilaton fields we cannot
adopt the free-like one. At present we do not find the proper normal ordering
yet. Here we only give a suggestion. To cancel the prefactor
$(\df^2 -\kappa)^{-1} $ of the Hamiltonian constraint, the most singular part
of the operator product between the two $\Pi_{\rho} $'s should behave like
$ \Pi_{\rho}(x) \Pi_{\rho}(y) \sim (\df^2 -\kappa)/ (x-y)^2 + \cdots $.
The similar structure should be realized for $\Pi_{\df} $.
The field dependence of the most singular term indicates that the theory has
the non-trivial coupling. This structure is very different from the other
quantum gravity models in two dimensions.

  After properly normal ordered, the commutation relations of the constraints
ought to satisfy the closed algebra without the central charge. Combining the
Hamiltonian and the momentum constraints as ${\hat T}_{\pm\pm}={1 \over 2}
(H \pm P) $, we get
$$
\eqalign{
         [{\hat T}_{\pm\pm}(x), {\hat T}_{\pm\pm}(y)]
          & = \pm 2i {\hat T}_{\pm\pm}(x) \delta^{\pp}(x-y)
               \pm i {\hat T}_{\pm\pm}^{\pp} \delta (x-y) ~,        \cr
         [{\hat T}_{++}(x), {\hat T}_{--}(y)] & =0 ~.               \cr
        } \eqno\eq
$$
This commutation relations generate the well-known Virasoro algebra without
central charge. This algebra guarantees the general covariance of the theory.

\chapter{{\bf Black hole dynamics}}

   Until now the arguments are completely non-perturbative. If we can solve
the physical state conditions exactly, the solution should include the
complete dynamics of black hole. Unfortunately it is a very difficult problem
so that we take an approximation. The original action (2.1) is order of
$1/\hbar $, but the Liouville part of ${\hat I}$ is zeroth order of $\hbar $.
However, if $\vert \kappa \vert $ is large enough, then it is meaningful to
consider the \lq\lq classical" dynamics of ${\hat I}$.  This is nothing but
the semi-classical approximation, which is valid only in the case of
$M \gg 1 $ and $N \gg 1$. In the other cases the quantum effect of gravitation
becomes important. The classical dynamics of ${\hat I}$ is ruled by the
equations ${\hat T}_{\alpha\beta} =0 $ and
the dilaton equation of motion
$$
\eqalign{
         -2 \pd_+ \df \pd_+ \df +2 \df \pd^2_+ \df
              -4 \df \pd_+ \df \pd_+ \rho
           & + {1 \over 2} \sum^N_{j=1} \pd_+ f_j \pd_+ f_j          \cr
            -\kappa &(\pd_+ \rho \pd_+ \rho -\pd^2_+ \rho +t_+ )=0 ~, \cr
         -2 \pd_- \df \pd_- \df +2 \df \pd^2_- \df
              -4 \df \pd_- \df \pd_- \rho
           & + {1 \over 2} \sum^N_{j=1} \pd_- f_j \pd_- f_j          \cr
            -\kappa & (\pd_- \rho \pd_- \rho -\pd^2_- \rho +t_- )=0 ~, \cr
         -2 \pd_+ \df \pd_- \df -2 \df \pd_+ \pd_- \df
           -\lambda^2 \df^2 \e^{2\rho} & -\kappa \pd_+ \pd_- \rho = 0  \cr
        }   \eqno\eq
$$
and
$$
        4 \pd_+ \pd_- \df +2\df \pd_+ \pd_- \rho
              +\lambda^2 \df \e^{2\rho} = 0  ~.
         \eqno\eq
$$
These are nothing but the CGHS equations\refmark{\cghs}
with the coefficient $\kappa$
instead of $N/12 $ in front of the Liouville part. Many authors have solved
these equations for $\kappa >0 $ and derived the dynamics of evaporating
black hole.\refmark{\rst,\w} Giving the expression (2.4) as the infalling
matter flux, we can get the exact solution of the equations along the line of
$x^+ =x^+_0 $
$$
      \pd_+ \df (x^+_0 ,x^- ) = {\lambda \over 2} \sqrt{-{x^- \over x^+_0 }}
               - {{M \over 2\lambda x^+_0 } \over
                      \sqrt{-\lambda^2 x^+_0 x^- - \kappa} } ~.
       \eqno\eq
$$
The (apparent) horizon, which is defined by the equation
$\pd_+ \df (x) =0 $,\refmark{\tih} locates at
$$
      x^- = - \sqrt{ \Big( {M \over \lambda^3 x^+_0 } \Bigr)^2 +
                     \Big( {\kappa \over 2\lambda^2 x^+_0 } \Bigr)^2  }
                 - {\kappa \over 2\lambda^2 x^+_0  } ~,
      \qquad x^+ = x^+_0    ~.
         \eqno\eq
$$
Initially the location of the horizon shifts to the outside of the classical
horizon defined through the solution (2.3) by quantum effects
(almost matter's effects). Then the black hole evaporates and the horizon
approaches to the singularity asymptotically.
The location of the singularity is determined by the equation
$\df^2 =\kappa $, which is easily proved by combining the equations (5.1) and
(5.2) properly (at $x^+ = x^+_0 $, it is
$x^{-} ={-\kappa \over \lambda^2 x^+_0 } $). It coincides with that determined
from the Hamiltonian constraint. Note that at the singularity the curvature is
singular, but the metric is regular.
As far as the gauge-fixed action is treated
classically, it seems that the horizon does not cross the singularity.
As mentioned before  the quantum mechanical region
$\kappa > \df^2 > 0 $ extends behind the singularity, where the quantum
gravitational effects become important.

   If $N $ is small, the non-anomalous quantum corrections of gravity
part maybe contribute to the dynamics and the approximation becomes bad.
Nevertheless we apply the approximation for $\kappa < 0 $
because we hope that some new
insights are obtained from the solution. If $\kappa < 0 $, the singularity
disappears. The location  of the horizon initially shifts to the inside of
the classical horizon. If the effective mass of the black hole  is defined
by $M_{BH} = \lambda \df^2 \vert_{\hbox{horizon}}$, this means that the
initial mass of the black hole is less than the infalling matter flux $M $.
After the black hole is formed, the positive flux comes in through the horizon
and the black hole mass increases. It seems that the horizon approaches to the
classical horizon asymptotically and becomes stable. If $\kappa =0 $,
the Liouville action disappears and the classical solution (2.3) is dominant.

   The problem of the information loss seems to come out in the case of
$\kappa > 0 $.
Then the black hole evaporates and the information seems to be lost.
However in this case the Liouville region extends behind the singularity.
So it appears that  there is a possibility that
the informations run away through this region by gravitational tunneling.
On the other hand, if $\kappa \leq 0 $, the Liouville region disappears.
But the black hole seems to be stable. In this case it appears that the
problem of the information loss does not exist.

\chapter{\bf Toward the quantization of spherically symmetric gravity}

   In this section we discuss the quantization of the spherically symmetric
gravitational system in 3+1 dimensions. If the 3+1 dimensional metric is
restricted as
$$
    \bigl( ds^{(4)} \bigr)^2 = g^{(4)}_{ab}dx^a dx^b =
          g_{\alpha\beta}dx^{\alpha} dx^{\beta} + G \df^2 d\Omega^2 ~.
     \eqno\eq
$$
where $d\Omega^2 $ is the volume element of a unit 2-sphere and $G $ is the
gravitational constant, the Einstein-Hilbert action becomes\refmark{\tih}
$$
\eqalign{
    I_{EH} &= {1 \over 16\pi G} \int d^4 x \hbox{$\sqrt{-g^{(4)}}$} R^{(4)} \cr
           &= {1 \over 4} \int d^2 x \hbox{$\sqrt{-g}$}
                \biggl(R_g \df^2 +2 g^{\alpha\beta}
                         \pd_{\alpha} \df \pd_{\beta} \df +{2 \over G}
                   \biggr) ~.                                              \cr
         }  \eqno\eq
$$
In the following we set $G=1 $.

   If the conformal matter defined by the action (2.1) is coupled and the
measures are defined by (2.6), the quantization is carried out in the
parallel with the case of the dilaton gravity. Then the gauge-fixed action
of the spherically symmetric gravity becomes
$$
     {\hat I}_{SSG} = {\kappa_s \over \pi} S_L (\rho, \bg)
                      + I_{EH} (\bgr,\df) + I_M (\bg,f) + I_{gh}(\bg,b,c) ~,
       \eqno\eq
$$
where the coefficient in front of the Liouville action
is\footnote\natural{This value is given by setting $\xi =1/2 $ in ref.7}
$$
       \kappa_s = {1 \over 12} (1-2+N-26)={N-27 \over 12} ~.
         \eqno\eq
$$
The nature of the quantum dynamics becomes the same as that of the dilaton
gravity. The differences are only quantitative.

   If both the black hole mass $M $ and the parameter $\kappa_s $ are large
enough, the classical dynamics of ${\hat I}_{SSG} $ is dominant.
This corresponds to taking the semi-classical approximation. As a classical
geometry we introduce the shock wave geometry similar to (2.3). It is given
by\refmark{\his}
$$
       ds^2 = - \biggl( 1- {2M \vartheta({\bar v}) \over r} \biggr)
                {{\bar u} \over {\bar u}+4M \vartheta({\bar v}) }
                 d{\bar u} d{\bar v}  ~,
           \qquad  \df = r ~,
          \eqno\eq
$$
This geometry is derived by
sewing the flat space time and the Schwarzshild black hole geometry along
the shock wave line. We first define that for $v < 0$ the metric is flat
$ds^2 = -du dv $, where $u=v-2r $, while for $v > 0$ the metric is the
Schwarzshild $ds^2 = -(1-{2M \over r}) du^{\star} dv $, where
$u^{\star}=v-2r^{\star} $ and $r^{\star}=r+2M \log ({r \over 2M}-1)$.
Next we relate the coordinate system $(r, v) $ with the coordinate
$({\bar u}, {\bar v})$ describing a gravitational collapse.
In the past infinity the geometry is asymptotically
flat so that we set ${\bar v}=v$ in the whole space time.
Let us take the metric
$ds^2 =-d{\bar u} d{\bar v} $ (or ${\bar u} =u $) for ${\bar v}<0 $. Then the
metric for ${\bar v} >0 $ is determined by the maching condition at
${\bar v}=0 $. The condition gives the relation
$d{\bar u} = du^{\star} ({\bar u}+4M)/{\bar u} $ and we get the
expression (6.5). This geometry is really a classical solution with the
infalling matter flux $T^f_{{\bar v}{\bar v}} = M\delta ({\bar v}) $.
In $({\bar u},{\bar v}) $ coordinate the location of the
horizon is given by ${\bar u} = -4M $.

   By substituting the classical shock wave geometry into the induced
energy-momentum tensor ${\hat T}^{\rho}_{{\bar u}{\bar u}} $ and transforming
it into that in the null coordinate
$u^{\star}$,\footnote\sharp{In $({\bar u}, {\bar v}) $ coordinate,
$t_{{\bar u}} $ and $t_{{\bar v}} $ of (3.8) vanish by the physical
requirement, but, in $(u^{\star}, v)$ coordinate, $t_{u^{\star}} $ is non-zero.
See the relation (3.7).}
we get the Hawking radiation\refmark{\dfu,\his,\l}
$$
      ({\hat T}^{\rho}_{u^{\star} u^{\star}}
             + {\kappa_s \over \pi} t_{u^{\star}} )
                     \vert_{v=+\infty \atop r: fixed}
             = {\kappa_s \over 64 \pi}{1 \over  M^2 }
                \biggl(1-{2M \over r} \biggr)^2
                 \biggl(1+ {4M \over r} + {12M^2 \over r^2} \biggr) ~.
        \eqno\eq
$$
In the spacial infinity $r \rightarrow \infty$, the fux becomes
$\kappa_s /64\pi M^2 $. This is  really the same as the result derived by
Hawking
if we replace $\kappa_s $ with $N/12 $.

   The quantum model of spherically symmetric gravity discussed above has
some problems. Here we adopt the conformal matter described by the
action (2.1). Strictly speaking, however, we should consider the action
such as $I_M = - \int d^2 x \sqrt{-g} \df^2 g^{\alpha\beta} \pd_{\alpha} f
\pd_{\beta} f$, which is derived by reducing the four dimensional action to
the two dimensional one. Ignoring $\df^2 $-factor corresponds to ignoring the
potential which appears when we rewrite the d'Alembertian in terms of the
spherical coordinate. The black hole dynamics is determined by the behavior
near the horizon so that it seems that this simplification does not change
the nature of dynamics.

   The other problem is in the definitions of measures. As the actions
are derived from the four dimensional ones, the two dimensional measures also
should be derived from the four dimensional one
$$
      <\delta g^{(4)},\delta g^{(4)}>_{g^{(4)}}
           = \int d^4 x \hbox{$\sqrt{-g^{(4)}}$} g^{(4)ab}g^{(4)cd}
           (\delta g^{(4)}_{ac}\delta g^{(4)}_{bd}
           + u \delta g^{(4)}_{ab}\delta g^{(4)}_{cd} ) ~,
       \eqno\eq
$$
where $u >0 $. From this definition we get
$$
\eqalign{
    &  < \vg, \vg >_g =
           \int d^2 x \dg \df^2 g^{\alpha \beta} g^{\gamma \delta}
            (\vg_{\alpha \gamma} \vg_{\beta \delta}
                + u \vg_{\alpha\beta} \vg_{\gamma\delta}) ~,        \cr
    &  < \vp, \vp>_g =
           \int d^2 x \dg \vp \vp  ~.                           \cr
        } \eqno\eq
$$
And also for the matter fields,
$$
      < \vf_j , \vf_j >_g =
           \int d^2 x \dg \df^2 \vf_j \vf_j
             \qquad (j=1, \cdots N)  ~.
       \eqno\eq
$$
The difference between (2.6) and (6.8-9) is apparent. The factor $\df^2 $ in
the measures of $g $ and $f $ prevents us from quantizing the spherically
symmetric gravity exactly. We expect that this factor also does not change the
nature of quantum dynamics drastically.

\ack{The authors would like to thank T. Yoneya for valuable
discussion. The work of K.H. is supported in part by Soryuushi Shogakukai.}

\appendix

   In this appendix we discuss the nature of the quantum dilaton gravity
defined by the action (2.8) and clarify the difference from ours.
The quantum theory of (2.8) is defined by
$$
      Z_{\chi}= \int {D_h (h)D_h (\chi)D_h (f)
                            \over {\rm Vol(Diff.)} }
                      \e^{iI_{\chi}(h,\chi,f)}       ~,
          \eqno\eq
$$
where $I_{\chi}=I_D(h, \chi)+I_M(h, f) $. The conformal gauge fixing is
carried out by separating the metric $h $ into the conformal factor $\rho $
and the background metric $\bg $ as $h =\bgr$. The $\rho $-dependence of the
measure of $\chi $ is evaluated as follows. Since the measure of $\chi $ is
invariant under the local shift, we can replace $\chi $ into
$\chi^{\pp}=\chi+{1 \over 2} \Delta_h^{-1} R_h $. Then the dilaton
action $I_D (h, \chi )$ becomes
$$
       I_D (h, \chi^{\pp}) = {1 \over 2\pi} \int d^2 x
                \hbox{$\sqrt{-h}$} \Bigl[
                  h^{\alpha\beta}\pd_{\alpha}\chi^{\pp}
                        \pd_{\beta}\chi^{\pp}
                  -{1 \over 4} R_h \Delta_h^{-1} R_h
                  +4\lambda^2 \exp \bigl( \chi^{\pp}
                           -{1 \over 2} \Delta_h^{-1} R_h \bigr)
                  \Bigr]  ~.
        \eqno\eq
$$
When in two dimensions the kinetic term of a field takes the
standard quardratic form and there is no derivative coupling, the short
distance behavior becomes that of the usual free field in two dimensions
since there is no divergence which could modify the singularity of free
field theory in perturbation expansion. Therefore the divergence structure
of $\chi^{\pp}$ field is the same as that of a single free boson.
This fact leads to the relation
$$
\eqalign{
         \int D_h (\chi)\e^{iI_D(h,\chi)}
          & = \int D_h (\chi^{\pp})\e^{iI_D (h,\chi^{\pp})}      \cr
          & = \exp \biggl[ i{1 \over 12\pi} S_L(\rho,\bg) \biggr]
               \int D_{\bg}(\chi^{\pp})\e^{iI_D(h,\chi^{\pp})}     \cr
          & = \exp \biggl[ i{1 \over 12\pi} S_L(\rho,\bg) \biggr]
               \int D_{\bg}(\chi)\e^{iI_D(h,\chi)}  ~.             \cr
        } \eqno\eq
$$
The relation for the matter and the ghost fields is given by (2.14).
According to the procedure of DDK, we finally get
$$
        Z_{\chi} = \int D_{\bg}(\Phi)\e^{i{\hat I}_{\chi}(\bg,\Phi)} ~,
          \eqno\eq
$$
where $\Phi $ denotes $\rho $, $\chi$, $f$, $b$ and $c$. The gauge fixed
action is
$$
\eqalign{
   {\hat I}_{\chi} = & {1 \over 2\pi} \int d^2 x \dbg \Bigl[
    {\hat g}^{\alpha \beta} \partial_{\alpha} \chi \partial_{\beta} \chi
    +2{\hat g}^{\alpha \beta}  \partial_{\alpha} \chi \partial_{\beta} \rho
       + \bcv \chi +4 \lambda^2  \e^{\chi+2\rho}          \cr
      & + \kappa_{\chi} (
         {\hat g}^{\alpha \beta} \partial_{\alpha} \rho \partial_{\beta} \rho
         + \bcv \rho )
       -{1 \over 2} \sum^N_{j=1}
         {\hat g}^{\alpha \beta} \partial_{\alpha} f_j \partial_{\beta} f_j
       \Bigr] +I_{gh}(\bg,b,c)                             \cr
        }  \eqno\eq
$$
with
$$
      \kappa_{\chi} ={1 \over 12}(1+1+N-26)={N-24 \over 12} ~.
        \eqno\eq
$$
Note that $\rho-\chi $ coupling including the derivative is the second order,
while the $\rho-\df $ coupling of the action (2.26) is the third order. This
difference is very important because the former becomes the free-like theory
after peforming the cannonical field transformation as mentioned below,
whereas  the latter is not so as mentioned in Sect.4.

  By defining the fields $X $ and $Y $ as \lq\lq linear" combinations of
$\rho $ and $\chi $\footnote\diamondsuit{Note that the Jacobian of the field
transformation is unity.}
$$
     \rho = X -{1 \over \kappa_{\chi}} Y ~, \qquad \chi = Y ~,
        \eqno\eq
$$
we can diagonalize the kinetic term of the gauge-fixed action
$$
\eqalign{
      {\hat I}_{\chi} = {1 \over 2\pi} \int & d^2 x \dbg \Bigl[
        \kappa_{\chi}{\hat g}^{\alpha \beta}
                \partial_{\alpha} X \partial_{\beta} X
       + \Bigl( 1- {1 \over \kappa_{\chi}} \Bigr)
             {\hat g}^{\alpha \beta}  \partial_{\alpha} Y \partial_{\beta} Y
       + \kappa_{\chi} \bcv X                                        \cr
       &  +4 \lambda^2  \e^{2 X + (1-{2 \over \kappa_{\chi}} ) Y }
       -{1 \over 2} \sum^N_{j=1}
         {\hat g}^{\alpha \beta} \partial_{\alpha} f_j \partial_{\beta} f_j
       \Bigr] +I_{gh}(\bg,b,c) ~.                                    \cr
        }  \eqno\eq
$$
Since there is no derivative coupling, the
short distance behavior of the diagonalized fields $X$ and $Y$ is that of the
usual free field in two dimensions. The diagonalized action is nothing but the
action derived by Bilal and Callan in ref.8.

\refout

\bye